\documentclass[sigconf]{acmart} 

\AtBeginDocument{%
  }

\setcopyright{acmcopyright}
\copyrightyear{2018}
\acmYear{2018}
\acmDOI{XXXXXXX.XXXXXXX}

\acmConference[Conference acronym 'XX]{Make sure to enter the correct
  conference title from your rights confirmation emai}{June 03--05,
  2018}{Woodstock, NY}
\acmPrice{15.00}
\acmISBN{978-1-4503-XXXX-X/18/06}

\usepackage[section]{easy-todo}

\usepackage{amssymb}
\usepackage{amsmath, amsfonts}
\usepackage{graphicx}
\usepackage{textcomp}
\usepackage{xcolor}
\usepackage{hyperref}
\usepackage{cleveref}
\usepackage{xspace}
\usepackage{paralist}
\usepackage{multirow}
\usepackage{subcaption}
\usepackage{tikz}
\usepackage{courier}
\usepackage{listings} 
\usepackage{balance}

\newcommand{\link}[1]{{\color{blue} #1}}
\newcommand{\cbert}{{\sc Code\-BERT}\xspace}
\newcommand{\cxglue}{{\sc Code\-XGLUE}\xspace}
\newcommand{\gcbert}{{\sc Graph\-Code\-BERT}\xspace}
\newcommand{\gcbsm}{{\sc GCB\-hybrid}\xspace}
\newcommand{\mlgcbert}{${\mathcal Polyglot}${\sc Graph\-Code\-BERT}\xspace}

\newcommand{\etal}{\emph{et al.}\xspace}
\newcommand{\ie}{\emph{i.e.},\xspace}
\newcommand{\eg}{\emph{e.g.},\xspace}

\definecolor{Gray}{gray}{0.3}
\tikzstyle{mybox} = [draw=black, very thick, rectangle, rounded corners, inner ysep=5pt, inner xsep=5pt, fill=gray!20]

\newcommand{\takeaway}[2]{
    \smallskip
    \noindent
    \begin{tikzpicture}
        \node [mybox] (box){%
        \centering
        \begin{minipage}{.465\textwidth}
        \fontsize{8.8}{10}\selectfont
        \textbf{Observation #1}. #2
        \end{minipage}
        };
    \end{tikzpicture}%
}

\begin{document}

\title{Learning code summarization from a small and local dataset}

\author{Toufique Ahmed}
\affiliation{%
  \institution{University of California, Davis}
  \city{Davis}
  \state{California}
  \country{USA}
  \postcode{95616}}
\email{tfahmed@ucdavis.edu}

\author{Premkumar Devanbu}
\affiliation{%
  \institution{University of California, Davis}
  \city{Davis}
  \state{California}
  \country{USA}
  \postcode{95616}}
\email{ptdevanbu@ucdavis.edu}

\renewcommand{\shortauthors}{Toufique Ahmed and Premkumar Devanbu}

\begin{abstract}

Foundation models (\eg  \cbert, \gcbert, CodeT5) work well for many software engineering tasks. These models are pre-trained (using self-supervision) with billions of code tokens, and then fine-tuned with hundreds of thousands of labeled examples, typically drawn from many projects. However, software phenomena can be very project-specific. Vocabulary, and other phenomena vary substantially with each project. Thus, training on project-specific data, and testing on the same project, is a promising idea. This hypothesis has to be evaluated carefully, e.g., in a time-series setting, to prevent training-test leakage. 
We compare several models and training approaches, including same-project training, cross-project training, training a model
especially designed to be sample efficient (and thus \emph{prima facie} well-suited for learning in a limited-sample same-project
setting) 
and a maximalist hybrid approach, fine-tuning first on many projects in many languages and then training on 
the same-project. We find that the maximalist hybrid setting provides consistent, 
substantial gains over the state-of-the-art, on many different projects in both Java and Python. 

\end{abstract}

%


\maketitle

\section{Introduction}
Machine learning applications in software engineering have been very successful in practice (\emph{e.g.,} Microsoft's CoPilot) and also on a wide range of more advanced applications~\cite{DBLP:journals/corr/abs-2102-04664}. Recently, there has been a great deal of interest in 
\emph{foundation models}~\cite{feng2020codebert,wang2021codet5,guo2020graphcodebert,kanade2020learning,ahmad-etal-2021-unified} which subjects  a very highly parametrized, high-capacity neural model
to a two-phase training regime. The first  unsupervised ``pre-training" phase is done with an enormous corpus, 
using a simple fill-in-the-blanks or predict-the-next-token/sentence regime. This phase can be
carried out on essentially any (unlabeled) code data harvested from the web. There is no task-specific goal here; the model simply  learns the statistics
of the input data.  The second phase, fine-tuning, trains on-task, and requires carefully curated, consistently labeled data, consisting typically of input-output pairs reflecting good, consistent, on-task performance. 

The challenges of creating  well-curated, de-duplicated, and yet relevant training datasets have been described by several authors~\cite{gros2020code,allamanis2019adverse}. Recent paper by Ahmed \emph{et al}~\cite{ahmed2021multilingual} and Chen \emph{et al}~\cite{chen2022transferability} explore some of the issues that arise with finding sufficient quantities of high-quality find-tuning data. Data availability may be limited because: \emph{first}, some languages (e.g., Ruby) are relatively less popular than other languages, and available high-quality data may be limited. \emph{Second}, the projects in a language may be skewed towards one application domain (e.g., Javascript for the web) and thus the performance of the trained model maybe somewhat uneven. 
\emph{Finally}, and most interestingly, when curating software engineering datasets, for on-task fine-tuning, yet another strange, unique, wrinkle arises:  \emph{project specificity}. 

It's well known that developers in different projects do behave somewhat  differently; they use different terminology, different algorithms, and
even different coding practices. 
As far back as  2009~\cite{zimmermann2009cross} it was observed that cross-project models don't perform as well
as in-project models on defect prediction tasks. These difficulties cross-over into language models;
even the earliest paper on language modeling for code~\cite{hindle2012naturalness} noted application-specific
effects. Subsequent work by Tu \emph{et al}~\cite{tu2014localness} and Hellendoorn~\cite{hellendoorn2017deep} noted the highly local, project- and even file-specific vocabularies
of source code, and proposed ways to handle them. 

This phenomenon offers an entirely new opportunity: \emph{Can project-specific training data improve performance?}  On the plus side, since vocabulary, coding styles etc are notoriously project-specific,  training and testing on the same project \emph{should} give better performance. This seems like an easy, low-hanging fruit. However, there are a couple of traps. First, when working within project, one has to be careful in partitioning training and test data, so that we only use
data that would be realistically available in practice. Second, within-project data may be quite substantially
limited in comparison to cross-project data. For this reason, within-project training regimes would require
models that learn well from fewer samples. 

For this reason, we believe that it would be useful to investigate approaches that would improve \underline{\emph{sample efficiency}} for the fine-tuning phase of foundation model training.  By ``improving sample efficiency" we mean the general goal of increasing the ability of machine-learning models to learn to perform better, with fewer training samples.  For example, a model $A$ that reliably performs as well as model $B$ with much fewer
training samples is a more ``sample efficient'' model. Sample efficient model $A$ both requires less data 
\emph{and} potentially trains much faster: thus saving human effort, time, and energy usage. 
Most of all, in settings where high quality training data is not
as abundant, model $A$ would be more attractive. 

Finally, as a sort of  ``stress-testing'' of the same-project tuning idea, we applied this to  an extremely well-tuned code summarization model,  to see if same-project training provided \emph{any improvement at all}. For this
we used the multilingual ``PolyGlot'' model, by 
 Ahmed \& Devanbu~\cite{ahmed2021multilingual}. They found that cross-project, multi-lingual training, using a very large fine-tuning set in many languages provided best-in-class performance (this ``PolyGlot" model was the chart-topper on the CodeXGlue Leaderboard\footnote{ See \url{https://microsoft.github.io/CodeXGLUE/}.} for a while, although CodeT5 has since reported better performance). We wondered whether even this extensively well-tuned model could be further improved on a specific project by further fine-tuning on the same project. One might expect that it wouldn't, since it is already so well trained\ldots but actually, it worked! 

In this paper 
we consider same-project fine-tuning, for the task of \emph{code summarization},  
and make the following contributions
\begin{enumerate}
\item
We investigate the benefits of within-project training, using a time-series scenario: we train only on ``past" data, and evaluate on ``future'' data. In the code summarization setting, this reflects a realistic setting where a developer asks for the summary of a piece of code, and we train only data already available that specific time point in the history of the project. We find that within-project training offers some advantages. 
\item
Second, we adapt the \gcbert foundation model specifically to improve its sample efficiency for code summarization; the resulting \gcbsm model, 
achieves high levels of sample-efficiency
and can  outperform the state of the art in some project-specific settings.
\item
We also found that the ``maximalist stress test'', adding project-specific fine-tuning to the already
extensively fine-tuned ``PolyGlot'' model, actually provides further benefits, and yields the best performance,
comfortably beating the state of the art CodeT5 model overall, with statistical \emph{and} practical significance (pairwise Wilcoxon test, with a difference in means, over all test samples, of about 3.7 BLEU-4; this is above the 2.0 BLEU-4 threshold difference that humans are
experimentally~\cite{roy2021reassessing} known to notice). 
\item
Finally, while ``PolyGlot''+same-project setting is most  performant, 
we do find that same-project training is remarkably efficient; even the largest projects use less than 2.5\% of
the time taken for cross-project training, while attaining comparable performance to current state-of-the-art 
\end{enumerate} 
The paper begins in the next section with some motivating explorations of project-specific phenomena relevant to the foundation model setting. Following that we present our methodology, followed by results, discussion, and related work. The paper ends with a brief speculation on future directions. 


\section{Background \& Motivation}
\label{back}

We begin with a brief overview of Foundation models~\cite{bommasani2021opportunities}, which are currently widely used in NLP and SE. 
Foundation models are trained in two stages (\ie pre-training and fine-tuning). In the pre-training stage, we train the models with billions of unsupervised tokens to teach the models the statistics of the language in a self-supervised way, using simple tasks like auto-regressively predicting the next token,
 filling in a blank in context, completing the next sentence, denoising an input, \emph{etc}.  
 These tasks are performed in a multi-layer deep network; the intermediate layers thus learn
 a representation (``embedding'') of the salient patterns of input token sequences in the code. 
Later we take the embeddings of the input sequence learned in the pre-training stage and further train it with supervised data in the fine-tuning stage. This pre-training+fine-tuning paradigm was first introduced in NLP by Devlin \etal~\cite{devlin2018bert}. They proposed an encoder-only model BERT, pre-trained with two training objectives: Mask Language Modeling (MLM) and Next Sentence Prediction (NSP). MLM is the most effective pre-training objective for encoder-only models where the model randomly masks out a certain percentage of tokens and unmasks them. Liu \etal showed that RoBERTA outperforms BERT using only MLM as a pre-training objective with some new training strategies (e.g., dynamic masking instead of static masking of a sequence)  and hyperparameters tuning~\cite{liu2019roberta}. BERT-style encoder-only models have inherent limitations for seq2seq generative tasks like Neural Machine Translations (NMT) because of the missing trained decoder. Two models, BART~\cite{lewis2019bart}  and T5~\cite{raffel2019exploring}   have well-trained decoders and perform well on seq2seq generative tasks.  

These models are not designed for code; subsequent research has refined these models (retaining the same basic scheme) for code and code-related natural language description. CodeBERT~\cite{feng2020codebert} and GraphCodeBERT~\cite{guo2020graphcodebert} are similar to the BERT model pre-trained with MLM and some code-specific pre-training objectives. PLBART~\cite{ahmad-etal-2021-unified} and CodeT5~\cite{wang2021codet5} are replications of BART~\cite{lewis2019bart} and T5~\cite{raffel2019exploring}  specially designed for SE tasks. Code-specific pre-trained models perform quite well on several SE tasks, including code summarization. The standard benchmark dataset \cxglue~\cite{DBLP:journals/corr/abs-2102-04664} is used to evaluate these models. The \cxglue includes a de-duplicated code summarization dataset prepared by modifying the CodeSearchNet~\cite{husain2019codesearchnet} dataset. \cxglue is a multilingual dataset (consisting of data from six languages) and has between 25K and  252K cross-project training samples for each language. Though some languages have relatively smaller samples (\ie Ruby and JavaScript), other  have very large training datasets (\ie Java and Python).

Ahmed and Devanbu~\cite{ahmed2021multilingual} have recently shown that multilingual training is beneficial for code summarization. Identifiers play a significant role in ML-based summarization, and they are mostly preserved across languages; this phenomenon enables cross-language training to work well. 
This prior work suggests that 
if methods from the same projects share similar identifiers, then same-project training can benefit the model. 
However, there are some issues when using same-project data. First, to be realistic, we can only use
data as it becomes available; thus at any point time, only \emph{past} data in the same project is available;
we cannot use data on classes, methods, \emph{etc} that haven't been created yet. Thus we perform
all our evaluations below in a ``time-series'' or ``time-partioned" setting. Second, and following from this
time-partitioned train-and-test approach, sample sizes get limited.  
Some times there is no more than a few hundred samples for each project, which differs greatly from the cross-project, cross-language setting where hundreds of thousands of instances can be  used to fine-tune the models. If the pre-trained models are sample-efficient, then same-project training can be proven effective for code summarization task. We will briefly look into the two preliminary, motivating questions (PMQs) :

\begin{description}

\item[PMQ 1] \emph{Do the different samples in the same project share more identifiers than samples drawn from  random projects?} If this were the case, one might hope that training with same project data would be especially advantageous. 
\item[PMQ 2]  \emph{Are the high capacity pre-trained models sample efficient? } If these models were not especially sample-efficient, then (because same-project data might be as abundant) we might have difficulty exploiting any same-project data synergies. 

\end{description}

\noindent{\underline{\em PMQ 1: Are identifiers preserved across same-project samples?} }
We conjecture that the same-project samples have higher identifier similarities because of domain and API usage similarity. Same-project samples also will use overlapping sets of user-defined class objects and identifiers. 
To evaluate this question, we perform a small experiment using  time partioned data. 
We take five projects from the Java \cxglue code summarization dataset, each with at least 200 samples and sort them according to their creation date. We perform the following steps.

\begin{enumerate}

\item Divide the first 200 samples of each project into two groups (\ie 1-100 and 101-200). 
\item Take each group and find the unique case-insensitive identifiers of the group. We repeat it for all five projects.
\item Take group I of project A and calculate the Jaccard index with group II of the same project.  
\item Now pair group I of project A with group II of all other projects and calculate the Jaccard index\footnote{Jaccard Index is calculated as  $\frac{\mid X \cap Y \mid}{ \mid X \cup Y \mid}$}.
\item Repeat steps 3 and 4 for all projects and observe the Jaccard Index difference.       

\end{enumerate}

Table~\ref{inter_intra} shows that we get the highest Jaccard indices (always 2-5 times higher than other positions) in the diagonal position where the data of both groups are coming from the same projects. Hence, same-project samples have higher identifier similarities.

\takeaway{1}{Same-projects samples likely to exhibit more identifier similarity than the cross-project samples. }

\noindent{\underline{\em PMQ 2: What is the fine-tuning sample efficiency of foundation models?} }
To observe the sample efficiency of foundations models, we consider two best performing models from each family of models (\ie, \gcbert from BERT-type encoder-only models and CodeT5 from seq2seq generative models). We also introduce a hybrid model \gcbsm in this paper (described in more
detail below), where we cascade the \gcbert encoder with a pre-trained decoder. 
For this experiment, we use the java \cxglue code summarization dataset. Note that CodeT5 is the best performing model for this task achieving 20.32 BLEU-4, where \gcbert reaches 19.22 BLEU-4 
We sample datasets of different sizes (10-300 examples) and observe the cross-project performance of the three models. Table~\ref{sample_efficiency} presents that with 300 code-project samples, CodeT5 achieves 18.23 BLEU-4 which is only about 2 BLEU-4 lower than what it performs with the complete dataset of ~165k samples. This, with about 550 times as much data! Therefore, we can conclude that models like CodeT5 are fairly sample-efficient and can perform well with a few data samples. However, CodeT5 struggles to summarize code well, when less than 150 training samples are available. 

In the same-project fine-tuning scenario, this situation is quite common, in many projects, as we argue
later.  
Happily, our  \gcbsm model is highly sample efficient,  and attains two-digit BLEU-4 even with ten examples. This is because \gcbsm's pre-trained decoder is especially trained to generate (denoised) comments. \gcbsm dominates the CodeT5 until 150 samples become available.

\begin{table}[h]
\centering
\resizebox{\columnwidth}{!}{%
\renewcommand{\arraystretch}{1.2}

\begin{tabular}{cccc}
\hline
\multicolumn{1}{c}{\#of samples}          & \multicolumn{1}{c}{\gcbert} & \multicolumn{1}{c}{\gcbsm} & \multicolumn{1}{c}{Codet5} \\ \hline
10                                          & 4.88                               & \textbf{11.37}                                & 1.38                        \\
50                                          & 9.29                               & \textbf{13.7}                                 & 1.84                        \\
100                                         & 10.02                              & \textbf{14.73}                                & 2.32                        \\
150                                         & 10.33                              & \textbf{14.98 }                               & 14.93                       \\
200                                         & 10.57                              & 15.51                                & \textbf{18.64}                       \\
250                                         & 10.73                              & 15.63                                & \textbf{18.81}                       \\
300                                         & 10.58                              & 15.71                                & \textbf{18.23 }                      \\  \hline
\multicolumn{1}{c}{Complete ($\sim$165k)} & \multicolumn{1}{c}{19.22}         & \multicolumn{1}{c}{19.97}           & \multicolumn{1}{c}{\textbf{20.32}}  \\ \hline
\end{tabular}
}
\vspace{0.05in}
\caption{Fine-tuning Sample-efficiency of  Foundation models }
\vspace{-0.2in}
\label{sample_efficiency}
\end{table}

\takeaway{2}{Pre-trained models can be adapted to be fine-tuning sample-efficient; such models are competitive with State-of-the-art for the code summarization task when samples are limited}

The following sections will discuss same-project training for code summarization using time series data and observe whether it can outperform the cross-project training performance with a few examples.

\begin{table*}[h]

\centering

\resizebox{\textwidth}{!}{%
\renewcommand{\arraystretch}{1.2}

\begin{tabular}{llccccc}
\hline
\multicolumn{2}{c}{\multirow{2}{*}{Projects}}                                                    & \multicolumn{5}{c}{Group II}                                                                                                                                                                                       \\ 
\multicolumn{2}{c}{}                                                                             & \multicolumn{1}{c}{oblac/jodd} & \multicolumn{1}{c}{wildfly/wildfly} & \multicolumn{1}{c}{orientechnologies/orientdb} & \multicolumn{1}{c}{Unidata/thredds} & \multicolumn{1}{c}{ngageoint/geopackage-android} \\ \hline
\multicolumn{1}{l}{\multirow{5}{*}{Group I}} & oblac/jodd                                        & \textbf{0.16}                            & 0.08                                 & 0.08                                            & 0.06                                 & 0.05                                              \\
\multicolumn{1}{l}{}                         & wildfly/wildfly                                   & 0.06                            & \textbf{0.16}                                 & 0.06                                            & 0.05                                 & 0.03                                              \\
\multicolumn{1}{l}{}                         & orientechnologies/orientdb                        & 0.07                            & 0.07                                 & \textbf{0.17 }                                           & 0.05                                 & 0.05                                              \\
\multicolumn{1}{l}{}                         & Unidata/thredds                                   & 0.07                            & 0.06                                 & 0.06                                            & \textbf{0.10                                } & 0.04                                              \\ 
\multicolumn{1}{l}{}                         & \multicolumn{1}{l}{ngageoint/geopackage-android} & \multicolumn{1}{c}{0.05}       & \multicolumn{1}{c}{0.04}            & \multicolumn{1}{c}{0.05}                       & \multicolumn{1}{c}{0.05}            & \multicolumn{1}{c}{\textbf{0.19}}                         \\ \hline
\end{tabular}
}
\vspace{0.05in}
\caption{Intra and inter project identifier overlap}
\vspace{-0.2in}
\label{inter_intra}
\end{table*}

\section{Methodology}
\label{method}
This section briefly describes the dataset preparation and foundation models we used for the evaluation.

\subsection{Dataset Preparation}
\label{datagen}
To evaluate the potential of same-project, sample-efficient training, we prepare a new dataset from \cxglue benchmark dataset. There are three reasons for choosing \cxglue

\begin{enumerate}

\item The dataset is known to be appropriately de-duplicated, thus avoiding issues raised in prior work~\cite{allamanis2019adverse,shia2022evaluation}
\item We can more easily baseline our approach; most foundation models have been evaluated on this dataset.
\item This dataset provides the complete path to all the functions with commit ID and line number. Using this information, we can find out the creation date of that particular function, and perform time-series
partitioning. 

\end{enumerate}

Preparing a same-project dataset, and then partitioning for training and test, has to be done carefully, to avoid 
the risk of possible data leakage from future to past instances during evaluation. Therefore, we perform time-series partitioning: we sort the samples from each project according to the creation date and perform an 80:20 split. 80\% of data are used for training, and later 20\% are randomly divided into test and validation sets. Note that we could not get the exact 80:20 splits for all projects because several functions share the same creation date at the point of splits. Now we will briefly discuss the process followed for assigning creation date to the examples and number of project used for the evaluation.    

\noindent{\underline{\em Assigning creation date to functions} }
As already mentioned, \cxglue provides the commit ID and path to the original function with specific line numbers. We use "git blame --ignore-rev" to extract the first commit date of a specific line. Note that most of the functions are multi-line, and we use start and end line numbers in the command to get the creation dates for all the lines. We consider the earliest date as the creation date of the complete function. We follow such a strategy because creation dates differ by line. Consider the following code snippet.

\begin{figure}[htb]
\captionsetup{aboveskip=-0pt,belowskip=0pt}
\vspace{-0.05in}
\centering
\begin{lstlisting}[language=Java,basicstyle=\scriptsize]
public static BeanCopy from(final Object source) {
	BeanCopy beanCopy = new BeanCopy(source);
	beanCopy.isSourceMap = source instanceof Map;
	return beanCopy;
}
\end{lstlisting}
\caption{{\em {\small Example for assigning creation date}}}
\label{fig-ex}
\end{figure}

For the example presented above (Fig.~\ref{fig-ex},) we get the same time-stamp (2015-08-26 12:57:28) for all the lines except the first one (2018-01-13 01:41:10). The approach we are following to extract the creation date has some limitations. ``git blame --ignore rev'' reports the earliest commit that changes a specific line. 
The first line was rewritten/edited from ``public static BeanCopy fromMap(Map source) \{'' to ``public static BeanCopy from(final Object source) \{'' on ``2018-01-13 01:41:10'', almost 2.5 years later than original function creation date (2015-08-26 12:57:28). We are interested in the original creation date instead of the last commit that changed a single program line. Considering the change made to the first line doesn't introduce any major change into the program. It does not even introduce or remove any identifier from the code. If we consider the times-tamps of all the function lines, that may help us get the original creation date. We consider the earliest commit time-stamp (2015-08-26 12:57:28) as the function creation date, for example presented in Fig.~\ref{fig-ex}.

There is one case when this approach will fail. If all the program lines have been modified over time at least once, we will fail to predict the actual creation date of the program because ``git blame --ignore rev'' will not be able to report the original creation date for any line of the program. However, this is very unlikely to happen. Another challenge is that we can only track down the history recorded on GitHub. Our approach will fail if the programs are created and edited on a local machine and then dumped into GitHub. However, we believe our system will still give a fair amount of time-sorted instances and can be used to evaluate the same-project data. Note that we are following a very conservative rule while splitting the project. Multiple functions with the same creation time-stamp will not appear on the training set and validation/test sets.

\noindent{\underline{\em Prepared datasets for different instance ranges} }
While creating the same-project dataset, we could only use the test data from \cxglue dataset. Pre-trained models (\ie, \gcbert, \gcbsm and CodeT5) are trained using the CodeSearchNet dataset, and there is a possibility of data leakage if we use the training set. We avoided the validation set because pre-trained models were evaluated on the validation set, and those models are likely to do well if the projects are taken from the validation set. We consider two popular programming languages (\ie Python and Java) for dataset generation.

Observing our experiments regarding sample efficiency, we divided our projects into three different training instance ranges.
\begin{enumerate}
\item Category I: Projects with 150+(more than 150) training samples (CodeT5 outperforms \gcbsm).
\item Category II: Projects with 100-150 training samples (\gcbsm performs well in that range).
\item Category III: Projects with 100-(less than 100) training samples (none of the models shows impressive performance).
\end{enumerate}

Table~\ref{dataset} presents the number of projects from each category. We selected 34 projects from 2 programming languages.

\begin{table}[h]
\centering
\resizebox{.5\columnwidth}{!}{%
\renewcommand{\arraystretch}{1.2}

\begin{tabular}{lcc}
\hline
\multicolumn{1}{c}{\multirow{2}{*}{Category}} & \multicolumn{2}{c}{Language}                           \\ 
\multicolumn{1}{c}{}                          & \multicolumn{1}{c}{Java} & \multicolumn{1}{c}{Python} \\ \hline
Category I                                      & 10                        & 7                           \\
Category II                                     & 6                         & 7                           \\
Category III                                    & 2                         & 2                           \\ \hline
\multicolumn{1}{l}{Total}                     & \multicolumn{1}{c}{18}   & \multicolumn{1}{c}{16}     \\ \hline
\end{tabular}
}
\vspace{0.05in}
\caption{Number of projects from each category}
\vspace{-0.2in}
\label{dataset}
\end{table}

\subsection{Foundation Models}

In this section, we briefly describe the foundation models that we use to compare the performance of cross-project and same-project training.  

\noindent{\underline{\em \gcbert} }
CodeBERT is one of the first BERT-type encoder-only pre-trained models specially designed for Software Engineering tasks. CodeBERT pre-trained with two objectives: i) MLM and ii) Replaced Token Detection (RTD). Though CodeBERT is successful in many downstream tasks, it does not use any code-related property. Guo \etal~\cite{guo2020graphcodebert}  describe an encoder-only foundation model, \gcbert, which uses two additional pre-training objectives (\ie edge prediction and node alignment) to MLM. The first additional task is to
 predict code structure edges, and the second aligns representations between source code and code structure. These two objectives incorporate data flow in the pre-training stage, a semantic-level code structure that encodes the relation of “where- the-value-comes-from” between variables. We evaluate \gcbert in this paper to evaluate the effectiveness of same-project training because \gcbert outperforms CodeBERT in all downstream tasks, including code summarization. 

\noindent{\underline{\em CodeT5} }
CodeT5~\cite{wang2021codet5} is a unified pre-trained encoder-decoder Transformer model well-suited for the seq2seq generative task. This model is pre-trained with three objectives: i) Masked Span Prediction (MSP), ii) Identifier Tagging (IT), and iii) Masked Identifier Prediction (MIP). CodeT5 learns improved embedding by leveraging the code semantics conveyed from the developer-assigned identifiers. It also achieves the state-of-the-art performance in \cxglue code summarization task. Both \gcbert and CodeT5 are primarily pre-trained 
with CodeSearchNet dataset. However, CodeT5 is pre-trained with some additional C and C\# datasets. 

\noindent{\underline{\em \gcbsm} }

\begin{figure}[h!]
    \centering
    \includegraphics[scale=0.30, trim={1.5cm 1cm 0cm 0cm}]{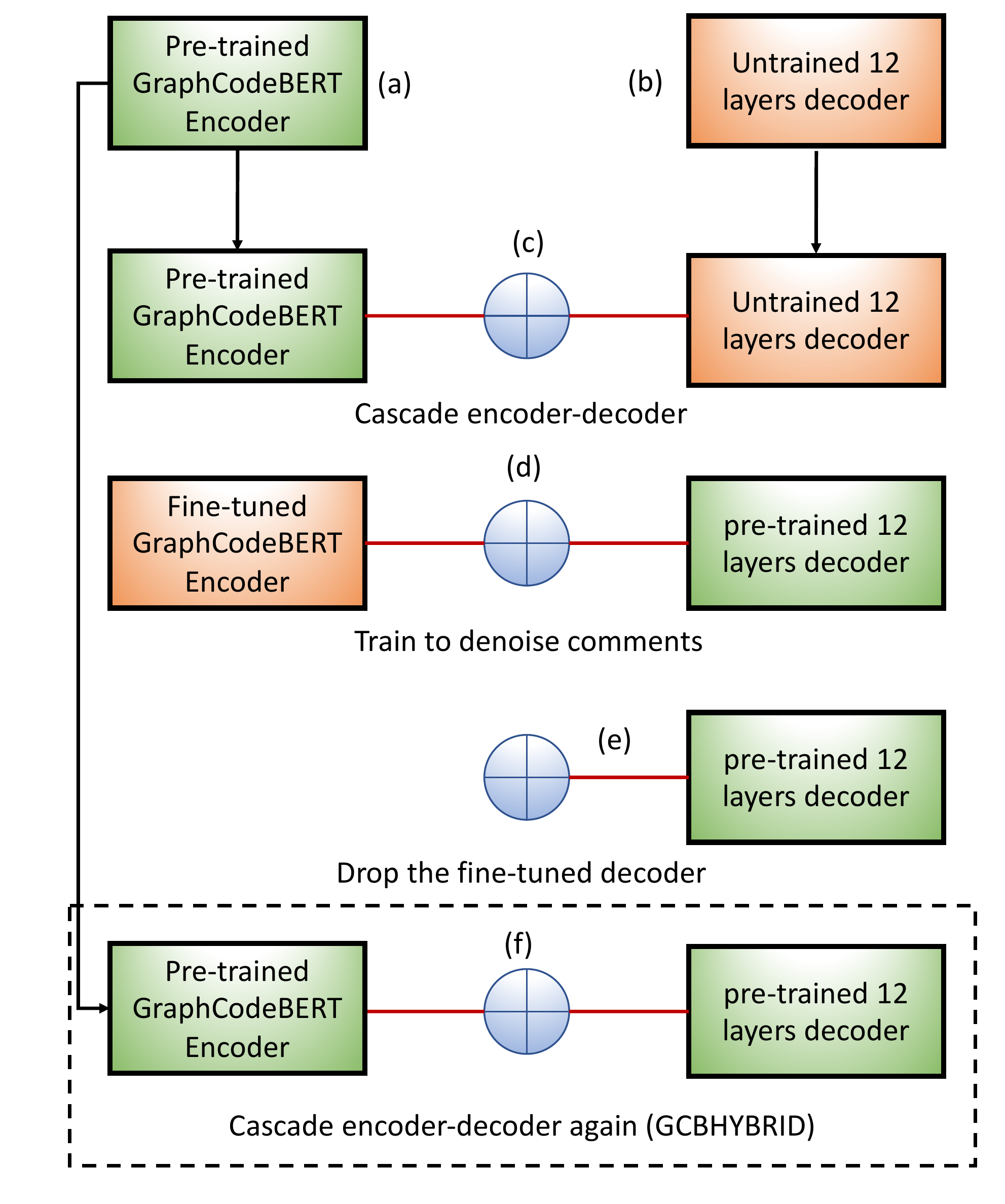}
    \caption{Steps for preparing \gcbsm }
    \label{gch}
\end{figure}

In the sample-efficiency experiment, presented earlier (\autoref{sample_efficiency}),  \gcbert reached only 10.58 BLEU-4 even after fine-tuning with 300 samples. Even 
the current SOTA, CodeT5, underperforms until we fine-tune with atleast 200 samples. However, both models do relatively well on the complete Java dataset, reaching 19.22 and 20.32 BLEU-4, respectively. Why do the pre-trained models underperform, with smaller fine-tuning datasets, even after training with billions of unsupervised tokens? \gcbert does not have a pre-trained decoder; it learns to generate comments only during fine-tuning. Thus, it cannot produce good comments until it's seen a large number of samples. On the other hand, CodeT5's pre-training also trains the decoder;  however, it's trained to ``refill' masked-out span of code, rather than a complete natural language description. The PLBART model is trained to denoise code and natural language description; but it failed to outperform \gcbert, even with its trained decoder, on Java code summarization (18.45 BLEU-4). 
We propose a hybrid model \gcbsm where we cascade the pre-trained \gcbert model with a specialized decoder pre-trained to denoise natural language description. Such a decoder help the model to do well on code summarization task by incorporating prior knowledge for generating natural language description.

Like \gcbert and CodeT5, We use CodeSearchNet dataset for training the decoder. We use only the given training partition of the CodeSearchNet to prevent any data leakage in the fine-tuning stage because our final test dataset is taken from the test partition of CodeSearchNet. We got approximately 2M natural language descriptions to proceed. Following BART, we implement five noising modes. Note that we apply two different types of noise modes to each sample to enhance the dataset. In the next segment of the paper, we will briefly explain the noise modes.

\noindent{\underline{\em Comment permutation} }      
With this noise mode, we take one code-related natural language description/comment at a time and shuffle the tokens in random order.

\noindent{\underline{\em Comment rotation} }
We randomly choose one token at a time and rotate the comment to bring that token to the first position of the statement. Repairing such noise helps to model to pick the starting of the comments.

\noindent{\underline{\em Token deletion} }
We randomly choose 15\% of the tokens and drop them from the comment. The model's task is to recover those dropped tokens and generate natural comments. 

\noindent{\underline{\em Token masking} }
Like token deletion, we randomly mask out 15\% of the tokens and ask the model to recover them and generate the comment using the decoder. Token masking is a comparatively easier task than token deletion. In token deletion, the model needs to learn both position and content of the missing token. 

\noindent{\underline{\em Token infilling} }
We select a random span with span lengths drawn from a Poisson distribution ($\lambda = 3$). We replace the span with a single token \textless{}mask\textgreater. The model will recover the complete missing span and learn about predicting the number of missing tokens in the span.

\begin{table}[h]
\centering
\resizebox{\columnwidth}{!}{%
\renewcommand{\arraystretch}{1.2}

\begin{tabular}{ll}
\hline
\multicolumn{1}{c}{Sequence Type}   & \multicolumn{1}{c}{Sequence}                                                                     \\ \hline
Original                              & Return next line with tag masked with whitespace .                                                \\
Comment permutation                  & with line . masked whitespace next tag with Return                                                \\
Comment rotation                     & masked with whitespace . Return next line with tag                                                \\
Token deletion                        & Return line tag masked with whitespace .                                                          \\
Token masking                         & Return next \textless{}mask\textgreater{} with \textless{}mask\textgreater{}  masked with whitespace . \\ 
\multicolumn{1}{l}{Token infilling} & \multicolumn{1}{l}{Return \textless{}mask\textgreater{}  tag masked with whitespace .}              \\ \hline
\end{tabular}

}
\vspace{0.05in}
\caption{Denoising natural language description}
\vspace{-0.2in}
\label{denoise}
\end{table}

Table~\ref{denoise} illustrates a comment mutated with the 5 noise modes. For training the decoder, we cascade a RoBERTa encoder with a newly created,
 12 layers transformer decoder model. To make this hybrid model work, we need to ensure both encoder and decoder use the same vocabulary. Therefore, we use the original \gcbert vocabulary  for this denoising task. To accelerate the training process, \gcbert was initialized with CodeBERT, and CodeBERT was loaded with the weights from the natural language RoBERTa. We also initialized our encoder with \gcbert and continued the denoising task for three epochs. Figure~\ref{gch} depicts the steps involved in preparing  \gcbsm. We startwith with (a) the pretrained \gcbert model and (b) untrained 12 layers transformer decoder.
These are adjoined (c)  together and trained together (d) for the de-noising task described above. 
After sufficient de-noising performance is achieved, the decoder has become
pretty good at generating comments. Now we detach just the decoder (e) and adjoin
it with the pre-trained original \gcbert encoder, to create the (f)  \gcbsm  model. 
We drop the fine-tuned encoder because it is subject to ``catastrophic forgetting'' of the knowledge learned in the base model~\cite{kirkpatrick2017overcoming}. 


\noindent{\underline{\em Stress test: the ``PolyGlot'' Model} } Finally, as a stress-test of the same-project  training approach, we wondered if even a very extensively fine-tuned model such as 
\mlgcbert~\cite{ahmed2021multilingual}, which fine-tuned on enormous (more than 900K) sample dataset, incorporate a diverse sample of project in many languages, could actually benefit
from fine-tuning on relatively small number of same-project samples. For this, we took the
published \mlgcbert model, and ran a few epochs of fine-tuning on same project data, and
evaluated in on same-project test data. We could do this without fear of training-test data overlap,
because the data partitions provided by \cxglue for pre-training and fine-tuning
guaranteed that this ``PolyGlot'' model had not seen these projects during its previous  pre-training and fine-tuning.

\begin{figure*}[h!]
    \centering
    \includegraphics[scale=0.30, trim={0 1cm 0cm 0cm}]{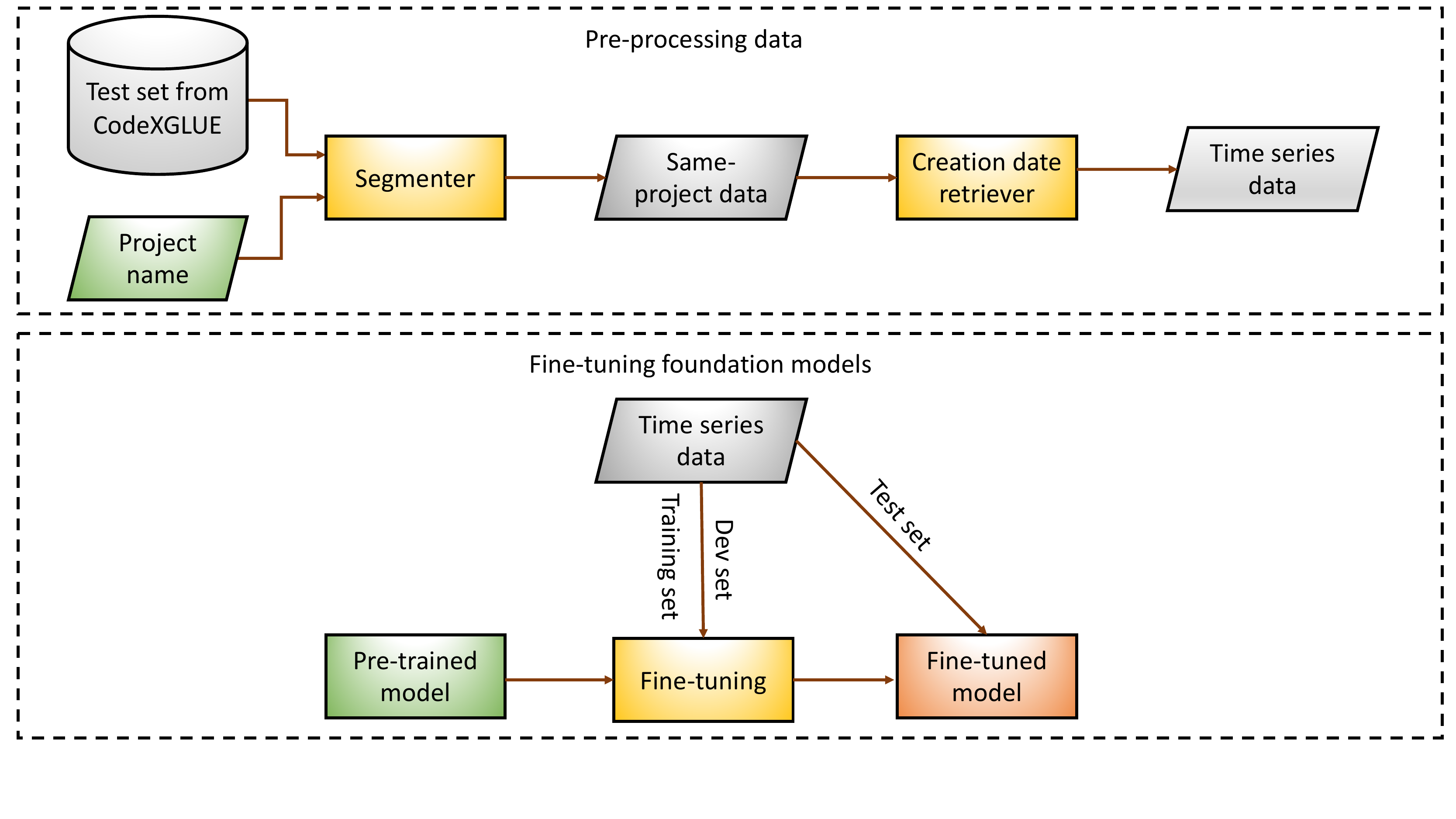}
    \caption{Complete pipeline for data generation and model training.}
    \label{pipeline}
\end{figure*}

\noindent{\underline{\em Complete pipeline and baselines} }
We are primarily investigating performance relative to the cross-project models, that are fine-tuned with hundreds of thousands of instances,  with same-project models fine-tuned only a few hundred samples. Figure~\ref{pipeline} presents the complete pipeline of our approach. In the pre-processing stage, we separate the same project data using a ``segmenter'' and convert it to time-series data following the approach described in Section~\ref{datagen} in ``creation date retriever'' stage. 
After preparing data, we fine-tune four foundation models (\ie \gcbert, \gcbsm, ``PolyGlot'', and CodeT5) for the code summarization task and compare them with the performance achieved by those three models in the cross-project setup (where there is no shortage of data).

\section{Results}

In this section, we evaluate same-project training for the code summarization task, in different settings.

\noindent{\underline{\em Fine-tuning  cross-project baselines \& same-project models} } 
As mentioned earlier (~\autoref{method}), we compare our approach with the models fine-tuned with abundant of cross-project data. We need to fine-tune the baseline models because we look for the BLEU-4 of a subset of test data while evaluating same-project training. The repository of the baseline models (\eg \gcbert\footnote{https://github.com/microsoft/CodeXGLUE/tree/main/Code-Text/code-to-text} and CodeT5\footnote{https://github.com/salesforce/CodeT5}) only provide cumulative (corpus) BLEU-4, 
which we cannot map into results on the (test) subset within the same project. 
For \gcbert, we fine-tune the model with 32 batch size, as recommended by \cxglue repository. However, we fine-tune the CodeT5 model with 24 batch size instead of 48 to fit the model into our Nvidia Titan RTX GPUs. We keep the other parameters unchanged. We also train our proposed \gcbsm model with the cross-project data for Java ($\approx 165k$ training samples) and Python ($\approx 251k$ training samples). For same-project training, we replace the cross-project samples with same-project data and fine-tune the models using the same codebases used for fine-tuning the baselines.

During fine-tuning, the cross-project models stop improving after 10 epochs, but same-project models continue improving even after 20 epochs, because of the smaller training set. Therefore, we fine-tune the same-project models for 30 epochs. Following all the code relevant foundation models~\cite{ahmad-etal-2021-unified,feng2020codebert,wang2021codet5}, we use smooth BLEU-4~\cite{lin2004orange} as the evaluation metric. 

\subsection{Effectiveness of same-project training on Category I projects}

Table~\ref{cat1java} presents the results of the Category I projects (with 150+ training samples) for Java. We achieve 18.65, 18.83, 19.52, and 19.72 on average with \gcbert, \gcbsm, CodeT5 and the "PolyGlot" models on the Java dataset in the cross-project fine-tuning setup. PolyGlot very slightly outperforms the other two models. In the same-project setup, we can see the encoder-only \gcbert lags, because the untrained decoder has too few samples from which to learn. However, same-project CodeT5 and \gcbsm perform really well, achieving 22.71 and 22.69  BLEU-4; the "PolyGlot" model excels, refinings it's already extensive multilingual
fine-tuning to reach 25.87 BLEU-4. 
All  models significantly improves over their cross-project counterpart (20.6\% for \gcbsm and 16.2\% for CodeT5).
 Roy \etal~\cite{roy2021reassessing} reported that less than 2 points do not guarantee systematic improvements in summarization quality and are not trustworthy as proxies of human evaluation. In this category, our best model shows over 6  BLEU-4 improvement over CodeT5 and the "PolyGlot" model,  which is 300\%  the Roy \etal threshold. Hence, same-project training introduces systematic improvement in code summarization task.

\begin{table*}[h!]

\centering

\resizebox{\textwidth}{!}{%
\renewcommand{\arraystretch}{1.2}

\begin{tabular}{lccccccccccc}
\hline
\multicolumn{1}{c}{\multirow{2}{*}{Projects}} & \multicolumn{3}{c}{Number of Samples}                                                      & \multicolumn{4}{c}{Cross-project}                                                                                                                                                               & \multicolumn{4}{c}{Same-project}                                                                                                                                                                \\ 
\multicolumn{1}{c}{}                          & \multicolumn{1}{c}{Training} & \multicolumn{1}{c}{Validation} & \multicolumn{1}{c}{Test} & \multicolumn{1}{c}{\gcbert} & \multicolumn{1}{c}{\gcbsm} & \multicolumn{1}{c}{CodeT5} & \multicolumn{1}{c}{\begin{tabular}[c]{@{}c@{}}PolyGlot \\ \gcbert\end{tabular}} & \multicolumn{1}{c}{\gcbert} & \multicolumn{1}{c}{\gcbsm} & \multicolumn{1}{c}{CodeT5} & \multicolumn{1}{c}{\begin{tabular}[c]{@{}c@{}}PolyGlot\\  \gcbert\end{tabular}} \\ \hline
oblac/jodd                                      & 913                           & 114                             & 114                       & 17.98                              & 17.93                                & 16.98                       & 17.63                                                                                  & 14.26                              & 20.54                                &\textbf{ 20.71}                       & 20.21                                                                                  \\
wildfly/wildfly                                 & 356                           & 44                              & 45                        & 12.53                              & 13.57                                & \textbf{16.16   }                    & 15.32                                                                                  & 10.41                              & 13.93                                & 14.67                       & 14.92                                                                                  \\
orientechnologies/orientdb                      & 346                           & 43                              & 44                        & 15.21                              & 14.21                                & 16.22                       & 14.51                                                                                  & 11.16                              & 16.05                                & 15.8                        & \textbf{16.52                                                                                 } \\
Unidata/thredds                                 & 1341                          & 167                             & 167                       & 14.18                              & 15.89                                & 16.36                       & 15.11                                                                                  & 11.82                              & 16.68                                & 16.07                       &\textbf{ 17.26                                                                                 } \\
ngageoint/geopackage-android                    & 239                           & 16                              & 16                        & 24.19                              & 22.91                                & 32.24                       & 21.42                                                                                  & 15.92                              & \textbf{40.72    }                            & 38.77                       & 34.95                                                                                  \\
RestComm/jain-slee                              & 184                           & 25                              & 25                        & 14.24                              & 16.21                                & 17.33                       & 12.87                                                                                  & 5.35                               & 13.20                                & \textbf{17.33                      } & 16.85                                                                                  \\
OpenEstate/OpenEstate-IO                        & 196                           & 10                              & 10                        & 21.89                              & 15.39                                & 18.7                        & 17.99                                                                                  & 12.44                              & 13.28                                & 3.01                        & \textbf{22.23                                                                                 } \\
tiefaces/TieFaces                               & 281                           & 14                              & 15                        & 36.74                              & 37.17                                & 32.37                       & 38.99                                                                                  & 25.53                              & 61.74                                & 64.02                       & \textbf{72.12                                                                                 } \\
jboss/jboss-common-core                         & 209                           & 17                              & 17                        & 17.16                              & 22.26                                & 17.33                       & 29.06                                                                                  & 11.01                              & 17.42                                & 21.83                       & \textbf{29.04                                                                                 } \\
rupertlssmith/lojix                             & 336                           & 41                              & 42                        & 12.36                              & 12.8                                 & 11.54                       & 14.32                                                                                  & 9.72                               & 13.5                                 & \textbf{14.65                      } & 14.62                                                                                  \\ \hline
\multicolumn{4}{l}{Average}                                                                                                                 & \multicolumn{1}{c}{18.65}         & \multicolumn{1}{c}{18.83}           & \multicolumn{1}{c}{19.52}  & \multicolumn{1}{c}{19.72}                                                             & \multicolumn{1}{c}{12.76}         & \multicolumn{1}{c}{22.71}           & \multicolumn{1}{c}{22.69}  & \multicolumn{1}{c}{\textbf{25.87}}                                                             \\ \hline
\end{tabular}
}
\vspace{0.05in}
\caption{Effectiveness of same-project fine-tuning for code summarization task on category I Java projects}
\vspace{-0.2in}
\label{cat1java}
\end{table*}

Same-project training also works for Category I python projects. As per Table~\ref{cat1python}, same-project 
"PolyGlot", on average, beats the next best prior work (CodeT5) by a solid 5.7 BLEU-4. Note that CodeT5
and \gcbsm also improve on same-project training, while \gcbert does not.

\begin{table*}[h!]

\centering

\resizebox{\textwidth}{!}{%
\renewcommand{\arraystretch}{1.2}

\begin{tabular}{lccccccccccc}
\hline
\multicolumn{1}{c}{\multirow{2}{*}{Projects}} & \multicolumn{3}{c}{Number of Samples}                                                      & \multicolumn{4}{c}{Cross-project}                                                                                                                                                               & \multicolumn{4}{c}{Same-project}                                                                                                                                                                \\ 
\multicolumn{1}{c}{}                          & \multicolumn{1}{c}{Training} & \multicolumn{1}{c}{Validation} & \multicolumn{1}{c}{Test} & \multicolumn{1}{c}{\gcbert} & \multicolumn{1}{c}{\gcbsm} & \multicolumn{1}{c}{CodeT5} & \multicolumn{1}{c}{\begin{tabular}[c]{@{}c@{}}PolyGlot \\ \gcbert\end{tabular}} & \multicolumn{1}{c}{\gcbert} & \multicolumn{1}{c}{\gcbsm} & \multicolumn{1}{c}{CodeT5} & \multicolumn{1}{c}{\begin{tabular}[c]{@{}c@{}}PolyGlot\\  \gcbert\end{tabular}} \\ \hline
apache/airflow              & 435 & 53 & 54 & 16.16 & 16.41 & 17.11 & 18.14 & 10.33 & 16.39 & 17.96 & \textbf{19.46 }\\
tensorflow/probability      & 425 & 53 & 53 & 17.9  & 18.86 & 22.18 & \textbf{22.39} & 13.49 & 18.19 & 21.33 & 20.76 \\
h2oai/h2o-3                 & 215 & 26 & 27 & 13.72 & 15.03 & \textbf{16.29} & 14.17 & 10.69 & 14.31 & 16.25 & 14.59 \\
Qiskit/qiskit-terra         & 376 & 43 & 43 & 23.13 & 22.81 & 24.27 & 19.93 & 17.65 & 21.24 & 24.13 & \textbf{25.65} \\
chaoss/grimoirelab-perceval & 188 & 24 & 24 & 14.59 & 11.69 & 14.79 & 11.12 & 26.26 & 38.7  & 36.82 &\textbf{ 50.27} \\
PyCQA/pylint                & 271 & 33 & 34 & 17.91 & 17.89 & 18.91 & 19.27 & 12.89 & 17.23 & \textbf{20.12} & 18.36 \\
SmokinCaterpillar/pypet     & 277 & 35 & 35 & 19.1  & 18.36 & 16.61 & 15.72 & 8.63  & 16.16 & 19.37 & \textbf{20.85} \\ \hline
\multicolumn{4}{l}{Average}                 & 17.50 & 17.29 & 18.59 & 17.25 & 14.28 & 20.32 & 22.28 &\textbf{ 24.28} \\ \hline
\end{tabular}
}
\vspace{0.05in}
\caption{Effectiveness of same-project fine-tuning for code summarization task on category I Python projects}
\vspace{-0.2in}
\label{cat1python}
\end{table*}

\takeaway{3}{Further same-project fine-tuning with 150+ samples helps the ``PolyGlot'' model  beat
all other prior conventionally fine-tuned models  by a substantial margin, on average.}

\subsection{Effectiveness of same-project training on Category II projects}

Table~\ref{cat2java} presents the results of the Category II projects (with 100-150 training samples) for Java. We measure 17.72, 17.83,  19.18, 18.93 BLEU-4 on average with \gcbert, \gcbsm, CodeT5, and "PolyGlot" models on the Java dataset in the  cross-project setting. The performance of the cross-project setting is consistent with the results we observe with Category I projects. However, CodeT5 underperforms with same-project training and scores only 5.33 BLEU-4 on average. In section~\ref{back}, we found that CodeT5 generally performs worse with less than 150 training samples. On the other hand, our decoder-enhanced model \gcbsm \emph{outperforms all the cross-project models} and achieves 21.34 BLEU -4, which is 2.16 higher than the best performing cross-models. The "PolyGlot" model surpasses even this model, reaching 23.39 BLEU, over
4.2 BLEU-4 better than the best conventionally fine-tuned (cross-project) model. 
 We find similar performance with Python also (Table~\ref{cat2python})

\takeaway{4}{With 100-150 same-projects samples, \gcbsm outperforms all the cross-project models. Again, "PolyGlot" does best overall. However, CodeT5 underperforms in this sample-range.}

\begin{table*}[h]

\centering

\resizebox{\textwidth}{!}{%
\renewcommand{\arraystretch}{1.2}

\begin{tabular}{lccccccccccc}
\hline
\multicolumn{1}{c}{\multirow{2}{*}{Projects}} & \multicolumn{3}{c}{Number of Samples}                                                      & \multicolumn{4}{c}{Cross-project}                                                                                                                                                               & \multicolumn{4}{c}{Same-project}                                                                                                                                                                \\ 
\multicolumn{1}{c}{}                          & \multicolumn{1}{c}{Training} & \multicolumn{1}{c}{Validation} & \multicolumn{1}{c}{Test} & \multicolumn{1}{c}{\gcbert} & \multicolumn{1}{c}{\gcbsm} & \multicolumn{1}{c}{CodeT5} & \multicolumn{1}{c}{\begin{tabular}[c]{@{}c@{}}PolyGlot \\ \gcbert\end{tabular}} & \multicolumn{1}{c}{\gcbert} & \multicolumn{1}{c}{\gcbsm} & \multicolumn{1}{c}{CodeT5} & \multicolumn{1}{c}{\begin{tabular}[c]{@{}c@{}}PolyGlot\\  \gcbert\end{tabular}} \\ \hline
real-logic/aeron                & 128 & 16 & 16 & 20.54                      & 19.58                            & 18.54                      & 17.84                       & 11.4                        & 23.22                      & 13.54                     & \textbf{27.07  }                          \\
boonproject/boon                & 123 & 14 & 15 & 20.14                      & 21.83                            & 25.16                      & 22.00                          & 16.94                       & 20.7                       & 1.87                      & \textbf{25.72}                            \\
Koekiebox-PTY-LTD/Fluid         & 100 & 13 & 13 & 24.46                      & 21.83                            & 22.51                      & 21.37                       & 20.1                        & \textbf{39.23}                      & 10.89                     & 31.6                             \\
lessthanoptimal/GeoRegression   & 120 & 15 & 16 & 16.91                      & 13.18                            & 16.35                      & 17.72                       & 8.75                        & 14.17                      & 3.64                      & \textbf{19.00 }                              \\
tony19/logback-android          & 118 & 16 & 16 & 12.15                      & 15.87                            & 18.88                      & 17.58                       & 6.61                        & 14.71                      & 1.5                       & \textbf{20.84 }                           \\
spring-projects/spring-security & 132 & 12 & 13 & 12.12                      & 14.73                            & 13.64                      & \textbf{17.04                      } & 5.53                        & 16.01                      & 0.54                      & 16.12                            \\ \hline
\multicolumn{4}{l}{Average}                   & \multicolumn{1}{c}{17.72} & \multicolumn{1}{c}{17.84} & \multicolumn{1}{c}{19.18} & \multicolumn{1}{c}{18.93} & \multicolumn{1}{c}{11.56} & \multicolumn{1}{c}{21.34} & \multicolumn{1}{c}{5.33} & \multicolumn{1}{c}{\textbf{23.39}} \\ \hline
\end{tabular}
}
\vspace{0.05in}
\caption{Effectiveness of same-project fine-tuning for code summarization task on category II Java projects}
\vspace{-0.2in}
\label{cat2java}
\end{table*}

\begin{table*}[h]

\centering

\resizebox{\textwidth}{!}{%
\renewcommand{\arraystretch}{1.2}

\begin{tabular}{lccccccccccc}
\hline
\multicolumn{1}{c}{\multirow{2}{*}{Projects}} & \multicolumn{3}{c}{Number of Samples}                                                      & \multicolumn{4}{c}{Cross-project}                                                                                                                                                               & \multicolumn{4}{c}{Same-project}                                                                                                                                                                \\ 
\multicolumn{1}{c}{}                          & \multicolumn{1}{c}{Training} & \multicolumn{1}{c}{Validation} & \multicolumn{1}{c}{Test} & \multicolumn{1}{c}{\gcbert} & \multicolumn{1}{c}{\gcbsm} & \multicolumn{1}{c}{CodeT5} & \multicolumn{1}{c}{\begin{tabular}[c]{@{}c@{}}PolyGlot \\ \gcbert\end{tabular}} & \multicolumn{1}{c}{\gcbert} & \multicolumn{1}{c}{\gcbsm} & \multicolumn{1}{c}{CodeT5} & \multicolumn{1}{c}{\begin{tabular}[c]{@{}c@{}}PolyGlot\\  \gcbert\end{tabular}} \\ \hline
Nic30/hwt               & 124 & 15 & 16 & 9.64                       & 15.81                      & 14.18                            & 16.93                            & 5.79                             & 12.8                       & 4.36                             & \textbf{17.05 }                     \\
vaexio/vaex             & 124 & 15 & 16 & 15.9                       & 17.23                      & 15.62                            & 17.08                            & 9.2                              & 13.6                       & 2.2                              & \textbf{18.19  }                    \\
assemblerflow/flowcraft & 113 & 14 & 15 & 14.01                      & 14.03                      & 14.39                            & 18.56                            & 9.04                             & 14.24                      & 1.1                              & \textbf{25.42}                      \\
funilrys/PyFunceble     & 104 & 13 & 13 & 16.63                      & 25.82                      & 22.97                            & 27.05                            & 19.11                            & 31.06                      & 5.67                             & \textbf{38.65 }                     \\
pyca/pyopenssl          & 100 & 13 & 13 & 19.81                      & \textbf{27.47}                      & 23.69                            & 23.1                             & 15.22                            & 24.87                      & 17.24                            & 25.4                       \\
LionelAuroux/pyrser     & 102 & 11 & 12 & 16.33                      & 16.91                      & 17.53                            & 13.75                            & 7.23                             & 13.98                      & 3.22                             & \textbf{20.79  }                    \\
OpenKMIP/PyKMIP         & 150 & 18 & 18 & 14.57                      & 15.66                      & 17.03                            & 18.8                             & 31.38                            & 39.11                      & 41.17                            & \textbf{42.59 }                     \\ \hline
\multicolumn{4}{l}{Average}           & \multicolumn{1}{c}{15.27} & \multicolumn{1}{c}{18.99} & \multicolumn{1}{c}{17.92} & \multicolumn{1}{c}{19.32} & \multicolumn{1}{c}{13.85} & \multicolumn{1}{c}{21.38} & \multicolumn{1}{c}{10.71} & \multicolumn{1}{c}{\textbf{26.87}} \\ \hline
\end{tabular}
}
\vspace{0.05in}
\caption{Effectiveness of same-project fine-tuning for code summarization task on category II Python projects}
\vspace{-0.2in}
\label{cat2python}
\end{table*}

\subsection{Effectiveness of same-project training on Category III projects}

Foundation models are pre-trained with billions of tokens. However, they are not trained to do code summarization task. They need enough samples to fine-tune the objective of the models. Though \gcbsm scored 11.37 BLEU-4 (\autoref{sample_efficiency}) with only 10 samples, it is much lower than we usually achieve with Cross-project models. Therefore, same-project training has some limitations. It requires a certain number of samples to achieve comparable results to cross-project models. We found that the models need at least 100 samples to compete with cross-project models from Category II and Category III projects. Table~\ref{cat3} has 4 projects in total (the first two from Python and later ones from Java). It shows that the same-project models are underperforming with less than 100 samples and achieve 14.99 with \gcbsm models. CodeT5 only achieves 2.84 BLEU-4, whereas all the cross-project models achieve more than 15.36 BLEU-4.    However, even in this case, ``PolyGlot" dominates after same-project fine-tuning, reaching 19.16;
however, in this setting, the improvement falls slightly shot of the 2 BLEU-4 threshold; however, it is important
to note that this approach could be used with other approaches to gain stronger improvements.

\begin{table*}[h]

\centering

\resizebox{\textwidth}{!}{%
\renewcommand{\arraystretch}{1.2}

\begin{tabular}{lccccccccccc}
\hline
\multicolumn{1}{c}{\multirow{2}{*}{Projects}} & \multicolumn{3}{c}{Number of Samples}                                                      & \multicolumn{4}{c}{Cross-project}                                                                                                                                                               & \multicolumn{4}{c}{Same-project}                                                                                                                                                                \\ 
\multicolumn{1}{c}{}                          & \multicolumn{1}{c}{Training} & \multicolumn{1}{c}{Validation} & \multicolumn{1}{c}{Test} & \multicolumn{1}{c}{\gcbert} & \multicolumn{1}{c}{\gcbsm} & \multicolumn{1}{c}{CodeT5} & \multicolumn{1}{c}{\begin{tabular}[c]{@{}c@{}}PolyGlot \\ \gcbert\end{tabular}} & \multicolumn{1}{c}{\gcbert} & \multicolumn{1}{c}{\gcbsm} & \multicolumn{1}{c}{CodeT5} & \multicolumn{1}{c}{\begin{tabular}[c]{@{}c@{}}PolyGlot\\  \gcbert\end{tabular}} \\ \hline
singularityhub/sregistry-cli & 98 & 10 & 11 & 10.98  & 12.64   & \textbf{14.6}    & 12.68 & 9.35   & 11.36   & 2.51 & 11.2   \\
deepmipt/DeepPavlov          & 89 & 14 & 14 & 15.27  & 16.8    & 16.72   & 18.39 & 8.4    & 20.79   & 5.69 & \textbf{26.5}   \\
apache/parquet-mr            & 84 & 10 & 11 & 18.01  & 17.51   & 19.5    & \textbf{21.83} & 10.16  & 14.15   & 1.2  & 21.57  \\
wro4j/wro4j                  & 94 & 13 & 14 & 17.12  & 15.92   & \textbf{17.99}   & 16.42 & 7.76   & 13.67   & 1.96 & 17.35 \\ \hline
\multicolumn{4}{l}{Average}                 & 15.35 & 15.72 & 17.20 & 17.33 & 8.92 & 14.99 & 2.84 & \textbf{19.16} \\ \hline
\end{tabular}
}
\vspace{0.05in}
\caption{Effectiveness of same-project fine-tuning for code summarization task on category III projects}
\vspace{-0.2in}
\label{cat3}
\end{table*}

\takeaway{5}{Same-project fine-tuning does help ``PolyGlot'' model dominate again; however, 
by itself it provides only modest gains with less than 100 samples, and
may have to be used together with other refinements}
\section{Discussion}

In this section, we will discuss the feasibility of applying same-project training and the computational cost needed for such training. We will also present one motivating example at the end of this section.

\subsection{Feasibility of same-project training}
\begin{figure}[h!]
    \centering
    \includegraphics[scale=0.40, trim={0 1cm 0cm 0cm}]{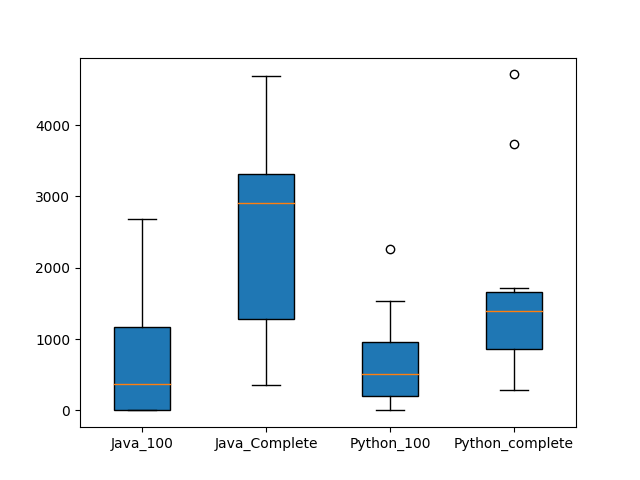}
    \caption{Complete lifespan and time needed to generate required number of samples.}
    \label{timespan}
\end{figure}

So far, we have discussed the possible benefits of same-project training;   higher BLEU-4 scores can
be attained with just 100 samples.  Our experiments suggest that an already extensively fine-tuned model like \mlgcbert can still benefit
from a few epochs same-project fine-tuning. 

However, how long must we wait, after a project starts,  to get 100 samples? This matters, because if it takes a long time to get enough samples, the benefit of same-project training is perhaps lower; one might just use cross-project training from older/existing projects. We look into the project-wise lifespans,  and time it takes to net 100 fine-tuning samples for our Category I and II projects. We track this data for both Python and Java, for each project in the CodeSearchNet data, from project inception
to the date the data was collected. Many of these projects are still active

Table~\ref{timespan} shows the distribution of the project time spans, in days, for both Java and Python projects. We show both the \emph{Complete} lifespan of the project (until last activity, or the current time, if still active and the time elapsed
until 100 fine-tuning  samples are available. 
It's evident that Java projects ``live'' longer than Python projects in our dataset. The median lifespan for java projects 2872 days (almost 8 years);  however, the time required to generate 100 fine-tuning samples is 335 days (less than 1 year). That means for nearly 7 years (85\% of the total lifespan), these projects could benefit from same-project fine-tuning. We have observed a similar situation for Python also. The median lifespan for Python projects 1365 days (almost 4 years), and the time required to generate 100 comments is 496 days (less than 1.5 years). Python projects could also benefit from same-project training for 64\% of their lifespan. Therefore, we can assume that the sample sizes sufficient for same-project training become available reasonably early and can be used for the remaining lifespan of the project.

\subsection{Computational cost of same-project training}

Same-project training is computationally much cheaper!
In fact, in our experiments same-project training run is persisted for three times as many epochs
as the cross-project run; we do this get better convergence with fewer samples; even so, we see significant
gains in fine-tuning costs.    
In the cross-project setting, we have 164,923 and 251,820 samples for Java and Python, respectively.  To compare with
the same-project setting, we choose the projects with the largest in-project fine-tuning datasets: 1341 for Java, and 435 for Python.  
Even these big projects, same-project training 
takes just 2.43\% (Java) and 0.52\% (Python) of the time of  cross-project training. Considering all the Java and Python Category I and II projects, the \emph{total} training samples will be 5,122 and 3,004, across all projects.. Cumulatively, \emph{for all projects}, same-project training  takes just 9.31\% and 3.57\% of the cross-project training time for Java and Python, respectively. Note that we compare the computational efficiency with respect to sample counts instead of time because the time required for training a certain number of samples varies due to server load.

\subsection{Motivating Example} 
Now we present one illustrative example where our same-project polyglot \gcbert model outperforms all the cross-project and same-project models. Table~\ref{exam3} presents the results from different models for the examples presented in Figure~\ref{exam1}. We can see cross-project \gcbert, \gcbsm, and polyglot \gcbert produce meaningful summaries for example I. 
Same-project \gcbsm also generates a complete sentence (0.36 BLEU-4). However, our polyglot \gcbert fine-tuned with the same project data gives the closest summary achieving 0.49 BLEU-4.

\begin{figure}[htb]
\captionsetup{aboveskip=-0pt,belowskip=0pt}
\vspace{-0.05in}
\centering
\begin{lstlisting}[language=Java,basicstyle=\scriptsize]


@Override
public void begin(InterpretationContext ic, String name, 
Attributes attributes) throws ActionException {
    hook = null;
    inError = false;
    String className = attributes.getValue(CLASS_ATTRIBUTE);
    if (OptionHelper.isEmpty(className)) {
        className = DefaultShutdownHook.class.getName();
        addInfo("Assuming className [" + className + "]");
    }
    try {
        addInfo("About to instantiate shutdown hook of type [" + 
            className + "]");
        hook = (ShutdownHookBase) OptionHelper.instantiateByClassName
            (className,ShutdownHookBase.class, context);
        hook.setContext(context);
        ic.pushObject(hook);
    }catch (Exception e) {
    	inError = true;
        addError("Could not create a shutdown hook of type [" + 
            className + "].", e);
        throw new ActionException(e);
    }
}




\end{lstlisting}
\caption{{\em {\small Code for motivating example}}}
\label{exam1}
\end{figure}

%
%
%
%
%
%

\begin{table*}[h]
\centering
\resizebox{.8\textwidth}{!}{%
\renewcommand{\arraystretch}{1.2}

\begin{tabular}{llllc}
\hline
\multicolumn{1}{c}{Original}                                                                                                     & \multicolumn{1}{c}{\begin{tabular}[c]{@{}c@{}}Cross-project or \\ same-project?\end{tabular}} & \multicolumn{1}{c}{Model name}                            & \multicolumn{1}{c}{Prediction}                                                                                                      & \multicolumn{1}{c}{BLEU-4} \\ \hline
\multirow{8}{*}{\begin{tabular}[c]{@{}l@{}}Instantiates a shutdown \\ hook of the given class \\ and sets its name .\end{tabular}} & \multirow{4}{*}{Cross-project}                                                                 & \gcbert                                                     & Initialize the shutdown hook .                                                                                                       & 0.11                        \\
                                                                                                                                   &                                                                                                & \gcbsm                                                      & Initialize the shutdown hook .                                                                                                       & 0.11                        \\
                                                                                                                                   &                                                                                                & CodeT5                                                     & \begin{tabular}[c]{@{}l@{}}This method is called at the \\ beginning of the action .\end{tabular}                                    & 0.13                        \\
                                                                                                                                   &                                                                                                & \begin{tabular}[c]{@{}l@{}}PolyGlot \\ \gcbert\end{tabular} & Parses a shutdown hook .                                                                                                             & 0.14                        \\ \hline
                                                                                                                                   & \multirow{4}{*}{Same-project}                                                                  & \gcbert                                                     & Get the the the .                                                                                                                    & 0.07                        \\
                                                                                                                                   &                                                                                                & \gcbsm                                                      & \begin{tabular}[c]{@{}l@{}}Instantiate a hook of \\ the given class .\end{tabular}                                                   & 0.36                        \\
                                                                                                                                   &                                                                                                & CodeT5                                                     & \begin{tabular}[c]{@{}l@{}}) \{ addInfo ( "Aboutto instantiate \\ shutdown hook oftype {[}" + ... \\ more random tokens\end{tabular} & 0.02                        \\
                                                                                                                                   &                                                                                                & \begin{tabular}[c]{@{}l@{}}PolyGlot \\ \gcbert\end{tabular} & \begin{tabular}[c]{@{}l@{}}Instantiates a new shutdown \\ hook of the given class .\end{tabular}                                     & 0.49   \\ \hline                     
\end{tabular}
}
\vspace{0.05in}
\caption{Predictions for the example presented in Fig.~\ref{exam1}}
\vspace{-0.2in}
\label{exam3}
\end{table*}


\section{Related Work}
\noindent{\underline{\em Code summarization} }
Code summarization is a widely studied problem in software engineering. 
Developers spend around 59\% of their time on activities somewhat relevant to program comprehension~\cite{xia2017measuring}, and good comments can ease the development and maintenance process by helping developers more quickly understand the meaning of code under maintenance~\cite{sridhara2010towards}. However, misaligned and outdated comments are prevalent in SE projects. Automatic code summarization can help provide more faithful \& current comments. Code summarization can also help write new comments.

We can closely relate the code summarization task to Neural Machine Translation (NMT) (\emph{e.g.,} translating English to German). In NMT, an encoder-decoder-based framework is used to do the translation task. Reachers in the SE domain have also adopted such a framework for code summarization tasks. Systems like CodeNN~\cite{iyer2016summarizing}, DeepCom~\cite{hu2018deep}, Astattgru~\cite{leclair2019neural}, Rencos~\cite{zhang2020retrieval}, NCS~\cite{ahmad2020summarization} and many more applied different kinds of deep learning architecture (\eg LSTM~\cite{sutskever2014sequence} and Transformers~\cite{vaswani2017attention}) on encoder-decoder framework and show good performance on code summarization task. Prior work~\cite{roy2021reassessing,shia2022evaluation,gros2020code} discuss the evaluation metrics and datasets that have been used for code summarization task.

\noindent{\underline{\em Foundation models for code summarization} }
Foundation models~\cite{feng2020codebert,ahmad-etal-2021-unified,qi2021prophetnet,phan2021cotext,liu2019roberta, mastropaolo2021studying,wang2021codet5} are currently state-of-the-art for the code summarization task.
Pre-training is key for foundation models, and helps them learn the language's statistical properties well. Since the model already knows about the language, a few examples are enough to train the model for a downstream task like code summarization. In this paper, we show that the pre-trained models are indeed sample-efficient and can outperform the models trained with cross-project data. Note that there are more than 30 papers that have been published in the last five years that follow some form of encoder-decoder architecture for code summarization~\cite{roy2021reassessing}. Comparing each model is beyond the scope of this paper. We primarily emphasize same-project, sample efficient fine-tuning that can be applicable to a wide range of models. Ahmed and Devanbu discuss augmenting the dataset using multilingual training and help models perform better. In contrast, we propose to reduce the sample count and perform well using same-project data. Autoregressive generative language models, such as GPT-3~\cite{brown2020language} have shown strong on-task performance, even after very limited fine-tuning; however, in our setting, without custom pre-training, as was done for \gcbert, and \gcbsm,
it's difficult to ensure that the (enormously sized) pre-training data used in these models was  not already pre-polluted with the data
we use for same-project testing; these enormously sized models are too costly to pre-train,
except for the  wealthiest organizations, so we omit these from our evaluation.

\section{Threats}
The main threats to our results arise from our evaluation approach. We explore some of them below

\noindent \paragraph{Data Pollution}
For external validity, and stability of results, 
it is important ensure that we \emph{never} test on data we used for pre-training or fine-tuning. 
The CodeXglue dataset for code summarization is split very carefully to avoid risk of data pollution; 
the pre-training data is separate from the fine-tuning data, and the test data is distinct from both of these. 
Our evaluation of the \gcbert and \gcbsm models adheres to this protocol.  The ``polyglot" model
is first fine-tuned on a large, multilingual dataset, of around a million samples, 
from CodeXGlue, before project-specific fine-tuning and same-project evaluation, on
held-out set of projects.

\paragraph{Data Duplication}
As described by Allamanis~\cite{allamanis2019adverse}, duplication lead to poor estimates of performance
that don't generalize. Fortunately CodeXglue~\cite{DBLP:journals/corr/abs-2102-04664} is very carefully de-duplicated,
and thus the performance numbers we report here can be expected to be fairly good. 

\paragraph{External Validity} 
External validity threats may arise from size of samples used to estimate performance, as well as whether
the representativeness of the samples. First,  our results have
statistical significance: we have compare the performance of our best model
(``polyglot'' model) with the SOTA (CodeT5) using a paired, non-parametric 2-sample test,
and can reject the null hypothesis that SOTA is the same or better than our best model. Second,
our average improvement on the BLEU-4 score is well above the 2 BLEU-4 threshold reported to be the barrier
for humans to detect. 

On the other hand, we have tested on a total of 18 Java projects and 16 Python projects. In almost every setting
our best model beats CodeT5; but in some cases, it does not. Therefore, some caution is warranted in assessing
the external validity of our results.

\section{Conclusion}
The existence, and impact, of Project-specific phenomena in software projects has been known for quite a while. The advent of
foundation models, which can be fine-tuned on-task, offers a possible direction to exploit project-specificity for better on-task performance. We explore this direction, for the code summarization task,  
with several popular foundation models.  We find that same-project training helps models exhibit competitive performance in several settings. In particular, we develop a new kind of \gcbert model, named \gcbsm, which combines
\gcbert with a specially trained decoder. \gcbsm exhibits very high sample-efficiency, which further enables exploitation of
project specificity; except ``polyglot'', \gcbsm does achieve state-of-the-art in some realistic same-project time-series settings. We also find that same-project
training offers substantial savings in computational cost. In addition to code summarization,  project-specific fine-tuning is a general idea that could well prove an useful adjunct for other tasks, such as defect prediction, fault localization, de-obfuscation, or automated patching. Finally, the same-project code summarization dataset and \gcbsm source code  
are made available anonymously at \link{\url{https://doi.org/10.5281/zenodo.6523229}}.

\balance
\bibliographystyle{ACM-Reference-Format}
\bibliography{acmart}
\end{document}